\documentclass[11pt]{article}
\usepackage{xcolor}
\usepackage{multirow}
\usepackage{enumerate}
\usepackage[scale=0.8]{geometry}

\usepackage{graphicx}
\usepackage[subrefformat=parens,labelformat=parens]{subfig}
\usepackage{hyperref}

\usepackage{subcaption}

\usepackage{amsfonts,amsmath,amssymb}
\usepackage{multirow,multicol}
\usepackage{bbm}
\usepackage{xspace}
\usepackage{slashed}

\usepackage[sort&compress, english]{cleveref}
\newcommand{\PP}{\mathbb{P}}

\newcommand{\cO}{\mathcal{O}}

\newcommand{\CY}{CY }

\newcommand{\FS}{\mathrm{FS}}
\newcommand{\cymetric}{\textsf{cymetric}}
\newcommand{\PySR}{\textsf{PySR}}

\usepackage[style=numeric-comp,url=true,doi=true,maxbibnames=99,sorting=none]{biblatex}

\bibliography{sample}

\numberwithin{equation}{section}

\usepackage{setspace}
\begin{document}
\phantom{}
\vspace{-.2cm}
\begin{center}
{

\LARGE {\bf Calabi--Yau Metrics with Full Moduli Dependence}\\[12pt]

\vspace{1cm}
\normalsize
{\bf{Andrei Constantin$^{a,b,}$}\footnote{a.constantin@bham.ac.uk}}
{\bf ,}
{\bf{Seung-Joo Lee$^{c,}$}\footnote{seungjoolee@yonsei.ac.kr}}
{\bf ,}
{\bf{Andre Lukas$^{b,}$}\footnote{andre.lukas@physics.ox.ac.uk}}
{\bf ,}
{\bf{Luca A. Nutricati$^{b,}$}\footnote{luca.nutricati@physics.ox.ac.uk}}
\bigskip}\\[0pt]
\vspace{0.23cm}
${}^a$ {\it 
School of Mathematics, University of Birmingham\\ Watson Building, Edgbaston, Birmingham B15 2TT, UK
}\\[2ex]
${}^b$ {\it 
Rudolf Peierls Centre for Theoretical Physics, University of Oxford\\
Parks Road, Oxford OX1 3PU, UK
}\\[2ex]
${}^c$ {\it 
Department of Physics, Yonsei University\\
Seoul 03722, Republic of Korea
}\\[2ex]
\end{center}
\vspace{15mm}

\begin{abstract}
Recent advances in numerical and machine-learning methods have enabled highly accurate constructions of Ricci-flat metrics on compact Calabi–Yau three-folds. For phenomenological applications it is crucial to understand how these metrics vary across moduli space. 
In this work, we construct approximate analytic expressions for Ricci-flat Calabi-Yau metrics with explicit complex-structure and K\"ahler moduli dependence by combining machine-learned numerical data with symbolic regression. Our approach is based on an explicit Ansatz for the K\"ahler potential with moduli-dependent coefficients. Fitting this Ansatz to numerical data and applying symbolic regression allows us to reconstruct analytic formulae for these coefficients, thereby obtaining approximate Ricci-flat metrics with explicit moduli dependence. We apply the construction to a one-parameter family of bi-cubic three-folds in $\mathbb{P}^2 \times \mathbb{P}^2$, achieving percent-level agreement with the underlying numerical data.
\end{abstract}

\clearpage
\section{Introduction}
One of the oldest ambitions of string phenomenology is to derive the couplings and masses of low-energy effective theories from the geometry and moduli of the extra dimensions rather than treating them as input data. Over the past few years substantial progress has been made in this direction. Numerical and machine-learning methods now make it possible to approximate Ricci-flat metrics on compact Calabi--Yau (CY) manifolds well enough to feed them into concrete physical calculations, including calculations of physical Yukawa couplings and fermion masses in heterotic compactifications \cite{Butbaia:2024tje,Constantin:2024yxh,Berglund:2024uqv}. 
Since these observables vary across moduli space, the numerical determination of Ricci-flat metrics must be repeated at many different points in moduli space.\\[2mm] 
Many quantities relevant for phenomenology are determined by topological or quasi-topological data. Physical couplings, however, also depend on wave-function normalisation factors and therefore require detailed metric information. Although these computations can now be carried out using numerical techniques~\cite{Donaldson:2005hvr, Douglas:2006rr, Braun:2007sn, Braun:2008jp, Headrick:2009jz,Cui:2019uhy, Ashmore:2019wzb, Anderson:2020hux,Jejjala:2020wcc,Ashmore:2021ohf,Larfors:2021pbb,Larfors:2022nep,Gerdes:2022nzr}, purely numerical representations have intrinsic limitations. Numerical methods, such as neural networks trained at a fixed point in moduli space, provide an approximation to the quantities of interest, but not explicit analytic expressions. As a result, important geometrical structures underlying the relation between moduli and low-energy physics may remain hidden. Moreover, numerical representations are difficult to extrapolate and cumbersome to differentiate with respect to the moduli in a controlled manner.\\[2mm] 
Despite their importance, explicit analytic Ricci-flat metrics are largely unavailable for compact Calabi--Yau three-folds. While remarkable constructions exist for K3 surfaces, where hyperk\"ahler geometry and instanton methods have led to analytic descriptions of Ricci-flat metrics~\cite{Kachru:2018van, Kachru:2020tat}, no comparable analytic description is currently known for compact Calabi--Yau three-folds.\\[2mm]
The purpose of this article is to combine and extend the approaches developed in Refs.~\cite{Lee:2025pue, Constantin:2026kmd}. These complementary works construct approximate Ricci-flat \CY metrics in analytic form with explicit dependence on complex-structure and K\"ahler moduli, respectively. The former focuses on complex-structure dependence, while the latter develops a framework that accommodates explicit K\"ahler-moduli dependence for manifolds with $h^{1,1}(X)>1$. Here we unify these approaches to construct approximate Ricci-flat metrics with simultaneous dependence on both complex-structure and K\"ahler moduli. Although the method is general, we illustrate it using a one-parameter family of bi-cubic three-fold hypersurfaces in $\mathbb{P}^2\times\mathbb{P}^2$.

\section{From numerical metrics to analytic formulae}
\label{Sec:general_approach}
Symbolic expressions for approximate Ricci-flat K\"ahler potentials on \CY manifolds with explicit moduli dependence have been constructed in Refs.~\cite{Lee:2025pue, Constantin:2026kmd}. Here we combine these approaches to obtain analytic approximations that depend simultaneously on both complex-structure and K\"ahler moduli. For mathematical background on K\"ahler geometry and manifolds with special holonomy see, for example, Ref.~\cite{joyce2000compact}.\\[2mm] 
Our starting point is an analytic Ansatz for the K\"ahler potential containing a finite set of free parameters. Ricci-flat metric data are generated numerically at a collection of points in the combined complex-structure and K\"ahler moduli space using the \cymetric~package~\cite{Larfors:2021pbb}. The parameters of the Ansatz are then determined by fitting to these data at each point in moduli space. Finally, symbolic regression is applied to the resulting parameter values in order to reconstruct their dependence on the moduli analytically.\\[2mm]
{\it Building a suitable Ansatz.} Before writing down the Ansatz, let us briefly  set up the basic framework and the notation. Consider a \CY hyper-surface $X$ embedded in an ambient space $\mathcal A$ which, for simplicity, we take to be a product of complex projective spaces\footnote{The method can be easily generalised to CY manifolds defined as hyper-surfaces or complete intersections in more general toric ambient spaces.}, $\mathcal A=\prod_{r=1}^{m}\mathbb P^{n_r}$ with homogenous coordinates
\begin{equation}
x = \left( x^{(1)}_1,\ldots,x^{(1)}_{n_1+1}, \ldots, x^{(m)}_1,\ldots,x^{(m)}_{n_m+1}\right)\; .
\end{equation}
K\"ahler classes $[J]$ on $X$ can be parametrised as
\begin{equation} 
[J] ~=~ \sum_{r=1}^{h^{1,1}(X)} t_r \,[J_r] ~\in~ H^{1,1}(X,\mathbb R)\, ,
\label{eq:Jexp}
\end{equation}
where the $[J_r]$ form a basis of $H^{1,1}(X,\mathbb R)$ and $t_r$ are the K\"ahler moduli, collectively denoted by $t=(t_1,\ldots,t_{h^{1,1}(X)})$. 
In the case of favourable embeddings, which will be our focus throughout this paper, $m=h^{1,1}(X)$ and the basis elements $[J_r]$ can be chosen as the restrictions of the ambient-space hyperplane classes.
Complex structure moduli will be denoted as $\psi = \left(\psi_1,\ldots, \psi_{h^{2,1}(X)}\right)$.\\[2mm]  
Following Ref.~\cite{Constantin:2026kmd} we write the Ricci-flat K\"ahler potential as a Fubini--Study piece plus a globally defined correction,
\begin{equation}
K (x,\psi,t) ~=~
K_{\FS}(x,\psi,t)\big|_{X} \,+\, \phi(x,\psi,t) 
\label{eq:general_ansatz}
\end{equation}
where the ambient-space Fubini--Study K\"ahler potential
\begin{equation}
K_{\FS}(x,\psi,t) ~=~
\sum_{r=1}^{m}\frac{t_r}{\pi}
\log\left(\sum_{a=0}^{n_r}|x_a^{(r)}|^2\right)
\label{eq:general_K_FS}
\end{equation}
has been restricted to the CY manifold. Although the complex-structure moduli do not appear explicitly in Eq.~\eqref{eq:general_K_FS}, they enter through the restriction to the hypersurface $X$. Since all evaluations are performed on sampled points of the \CY manifold, this implicit dependence presents no practical difficulties.\\[2mm]
The Ansatz for the globally defined function $\phi$ is built from a finite-dimensional space of sections of an ample line bundle $L = \mathcal{O}_X(k)$, where $k = (k_1, \dots, k_m)$, and $(s_I)_{I=1,\dots,h^0(X,L)}$ denotes a basis of global holomorphic sections of $L$. Specifically, the Ansatz reads
\begin{equation}
\begin{aligned}
&\phi\left(x,\psi,t\right) ~=~ V(t)^{1/3}\,\frac{\sum_{I,J} \alpha_{IJ}(\psi, t_1/t_m,\ldots, t_{m-1}/t_m)\,s_I(x)\,\overline{s_J(x)}} {\prod_{r=1}^{m} \left(\sum_{a=0}^{n_r} |x^{(r)}_a|^2\right)^{k_r}} \, ,
\end{aligned}
\label{eq:general_phi}
\end{equation}
where the \CY volume factor $V(t)= \frac{1}{3!}\int_X J^3$ has been introduced to ensure the correct scaling of the metric under a rescaling of the K\"ahler parameters $t_r \to \lambda t_r$.\\[2mm] 
Generalising the construction of Ref.~\cite{Constantin:2026kmd}, the coefficients $\alpha_{IJ}$ are taken to depend on both the complex-structure moduli $\psi$ and the K\"ahler moduli ratios $t_i/t_m$ for $1\leq i\leq m-1$. Since $\phi$ is globally defined, it does not modify the K\"ahler class, so the K\"ahler class associated to $K$ is entirely determined by the Fubini--Study contribution and is parametrised by Eq.~\eqref{eq:Jexp}. For each point in moduli space, the coefficients $\alpha_{IJ}$ will be determined by fitting the Ansatz to numerical Ricci-flat metric data obtained from machine learning. As usual, increasing the degree $k$ enlarges the space of sections and is expected to improve the accuracy of the Ansatz.\\[2mm] 
{\it The role of discrete symmetries.} In the following analysis we require that the \CY hypersurfaces under consideration admit a large discrete symmetry. This symmetry is reflected at the level of the metric and the K\"ahler potential and is used in order to simplify the structure of the Ansatz. (See Refs.~\cite{Lee:2025pue, Constantin:2026kmd} for an argument establishing the fact that a holomorphic automorphism of a CY manifold is respected by the Ricci-flat metric.)\\[2mm] 
{\it Retrieving analytic results.} 
Given the analytic Ansatz discussed above, we first use neural networks to learn the Ricci-flat K\"ahler potential numerically on a grid in moduli space and determine the corresponding Ansatz coefficients by fitting to the learned data. We then apply symbolic regression, implemented with \PySR~\cite{PySR}, to reconstruct explicit moduli-dependent expressions for these coefficients. This yields an approximate analytic formula for the Ricci-flat K\"ahler potential with explicit full moduli dependence.

\section{The bi-cubic three-fold}
\subsection{Geometry and conventions}
To set the stage, let us consider bi-cubic three-folds $X$ defined as hypersurfaces in the ambient space $\PP^2\times\PP^2$. We denote the homogeneous coordinates on the two projective factors by $x_a$ and $y_a$, respectively, where $a=0,1,2$. On the patch $x_0\neq0$, $y_0\neq0$, one may use the affine coordinates $x_a/x_0$ and $y_a/y_0$ with $a=1,2$. The manifold $X$ is defined as the zero locus of a polynomial $P$ of bi-degree $(3,3)$ and has Hodge numbers
\begin{equation*}
(h^{1,1}(X),h^{2,1}(X))=(2,83)\,.
\end{equation*}
The K\"ahler class can be written as
\begin{equation*}
[J]=t_1[J_1]+t_2[J_2]\,,
\end{equation*}
where $[J_1]$ and $[J_2]$ are the restrictions to $X$ of the Fubini--Study classes on the two ambient projective-space factors. Relative to this basis, the K\"ahler cone is characterised by $t_1,t_2>0$, and the volume is
\begin{equation*}
V(t_1,t_2)=\frac{3}{2}\,t_1t_2(t_1+t_2)\,.
\end{equation*}
There exist families of bi-cubics with large discrete symmetry groups that considerably simplify the analysis. We now describe one such family.

\subsection{Symmetry-reduced Ansatz}
The generators of the symmetry group under consideration act linearly on the homogeneous coordinates as
\begin{equation} \begin{array}{rclcl} (\mathbb{Z}_3^{(x)})_a ~\!\!\!\!&:& x_a\mapsto \omega x_a&,& x_b\mapsto x_b\;\;\mbox{for}\;\; b\neq a\,,\;\; y_c\mapsto y_c\;\;\mbox{for}\;\; c=0,1,2\\[2mm] (\mathbb{Z}_3^{(y)})_a ~\!\!\!\!&:& y_a\mapsto \omega y_a&,& y_b\mapsto y_b\;\;\mbox{for}\;\; b\neq a\,,\;\; x_c\mapsto x_c\;\;\mbox{for}\;\; c=0,1,2\\[2mm] S_3^{(x,y)}&:& x_a\mapsto x_{\sigma(a)}&,& y_a\mapsto y_{\sigma(a)}\;\;\mbox{where}\;\;\sigma\in S_3\\[2mm] \mathbb{Z}_2^{(x,y)}&:& x_a\mapsto y_a&,& y_a\mapsto x_a\;\;\mbox{for}\;\; a = 0,1,2 \end{array} 
\label{eq:symmetry}
\end{equation}
where $\omega = e^{2\pi i/3}$. Note that the generators in the first three lines act trivially on the K\"ahler moduli, while the last one exchanges the two moduli, $t_1\leftrightarrow t_2$. 
To obtain the induced projective symmetry group, one must quotient by the two diagonal $\mathbb Z_3$ actions
$x_a\mapsto\omega x_a$, $y_a\mapsto y_a$
and
$x_a\mapsto x_a$, $y_a\mapsto\omega y_a$, with $a=0,1,2$,
which are part of the projective rescalings of the two ambient $\PP^2$ factors. The resulting projective symmetry group is
\begin{equation}
G \cong \Bigl((\mathbb Z_3^{(x)})^2 \times (\mathbb Z_3^{(y)})^2\Bigr)\rtimes (S_3\times \mathbb Z_2)~,
\end{equation}
of order $|G|=3^4\cdot 6\cdot 2 = 972$.
The most general bi-cubic defining polynomial invariant under this symmetry group is
\begin{equation}
P=P_1+\psi P_2,\qquad
P_1=\sum_a x_a^3y_a^3,\qquad
P_2=\sum_{a\neq b}x_a^3y_b^3,
\label{eq:def_P}
\end{equation}
where $\psi$ is the single remaining complex-structure modulus.\\[2mm]
The bi-cubic three-folds defined as the zero locus of $P$ become singular when $\psi = -\frac12,0,\pm 1, \infty$ and are smooth otherwise. Of the five values of the parameter, $\psi=-\frac{1}{2}, -1$ and $\infty$ each leads to a finite number of point-like singularities on the bi-cubic hypersurface, while $\psi=0$ and $\psi=1$ correspond to severer singular loci of complex dimensions one and two, respectively. \\[2mm]
Our next step is to specialise the K\"ahler potential Ansatz in Eq.~\eqref{eq:general_ansatz} to the above one-parameter family of bi-cubics. We choose the line bundle $L=\cO_X(2,2)$ whose sections form a space of dimension $h^0(X,L)=36$. This means that the hermitian matrix $\alpha_{IJ}$ in the Ansatz~\eqref{eq:general_ansatz} is of size $36\times 36$ and, hence, contains $1296$ real parameters. Consequently, for a general bi-cubic this is a rather complicated problem. However, for the one-parameter family in Eq.~\eqref{eq:def_P} the sections combine into only eight quantities $I_i$, explicitly defined below, with well-defined transformation properties under $G$. In terms of these quantities, the Ansatz~\eqref{eq:general_ansatz} takes the form 
\begin{equation}
\phi(x,\psi,t_1,t_2) ~=~ V(t_1,t_2)^{1/3}
\sum_{i=0}^{7}\alpha_i\!\left(\psi,\frac{t_1}{t_2}\right)
\frac{I_i(x,y)}{\bigl(\sum_A|x_A|^2\bigr)^2\bigl(\sum_A|y_A|^2\bigr)^2}\,.
\label{eq:bicubicphiAnsatz}
\end{equation}
The coefficients $\alpha_i$ depend on the complex structure parameter $\psi$ and the one ratio, $t_1/t_2$, of K\"ahler parameters and will be determined by our fit to the data.
The factor $V^{1/3}$ ensures the correct homogeneity under an overall rescaling of the K\"ahler class. The quantities $I_i$ are invariants of the subgroup of $G$ that preserves the K\"ahler structure. This subgroup is generated by all transformations in Eq.~\eqref{eq:symmetry} except for the $\mathbb{Z}_2^{(x,y)}$ symmetry in the last line, which exchanges the two K\"ahler parameters $t_1$ and $t_2$. Explicitly, the invariants are given by
\begin{align}
I_0 &= \sum_{a<b}|x_a|^2|x_b|^2|y_a|^2|y_b|^2\,,
&
I_1 &= \sum_{a\neq b,c\atop b<c}|x_a|^4|y_b|^2|y_c|^2\,,
\notag\\
I_2 &= \sum_{a\neq b}|x_a|^4|y_a|^2|y_b|^2\,,
&
I_3 &= \sum_{a\neq b}|x_a|^4|y_b|^4\,,
\notag\\
I_4 &= \sum_a |x_a|^4|y_a|^4\,,
&
I_5 &= \sum_{a,b,c\,\mathrm{distinct}}|x_a|^2|y_a|^2|x_b|^2|y_c|^2\,,
\notag\\
I_6 &= \sum_{a\neq b}|y_a|^4|x_a|^2|x_b|^2\,,
&
I_7 &= \sum_{a\neq b,c\atop b<c}|y_a|^4|x_b|^2|x_c|^2\,.
\label{eq:bicubicKInvariants}
\end{align}
The action of $\mathbb{Z}_2^{(x,y)}$ leaves $I_i$ invariant for $i=0,3,4,5$, while it exchanges $I_1\leftrightarrow I_7$ and $I_2\leftrightarrow I_6$. Consequently, the coefficients $\alpha_i$ satisfy
\begin{align}
&\alpha_1\!\left(\psi,t_{12}\right)=\alpha_7\!\left(\psi,t_{21}\right)\, ,
\qquad
\alpha_2\!\left(\psi,t_{12}\right)=\alpha_6\!\left(\psi,t_{21}\right)\, , \nonumber \\[5pt]
& \qquad ~~ \alpha_i\!\left(\psi,t_{12}\right)=\alpha_i\!\left(\psi,t_{21}\right)\,,
\ \, {\rm for~} i=0,3,4,5\,,
\label{eq:alphasymmetry}
\end{align}
where $t_{12}\equiv t_1/t_2=t_{21}^{-1}$. These relations provide non-trivial consistency checks on the coefficients extracted from the numerical data.

\subsection{Results}
\begin{figure}[h!]
    \centering
    \begin{minipage}{0.45\textwidth}
        \centering
        \includegraphics[width=\linewidth]{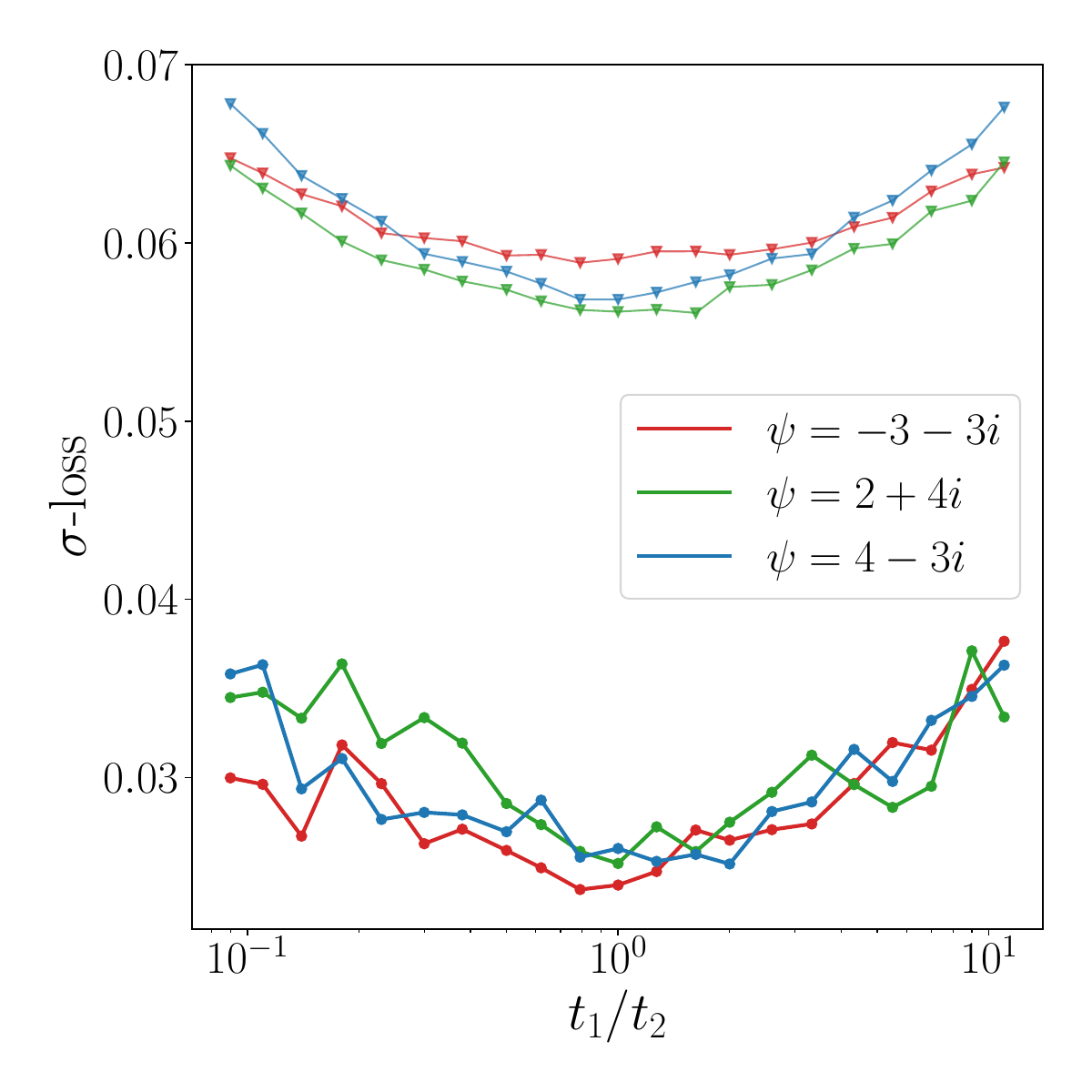}
    \end{minipage}
    \begin{minipage}{0.47\textwidth}
        \centering
        \vspace{-0.6cm}
        \includegraphics[width=\linewidth]{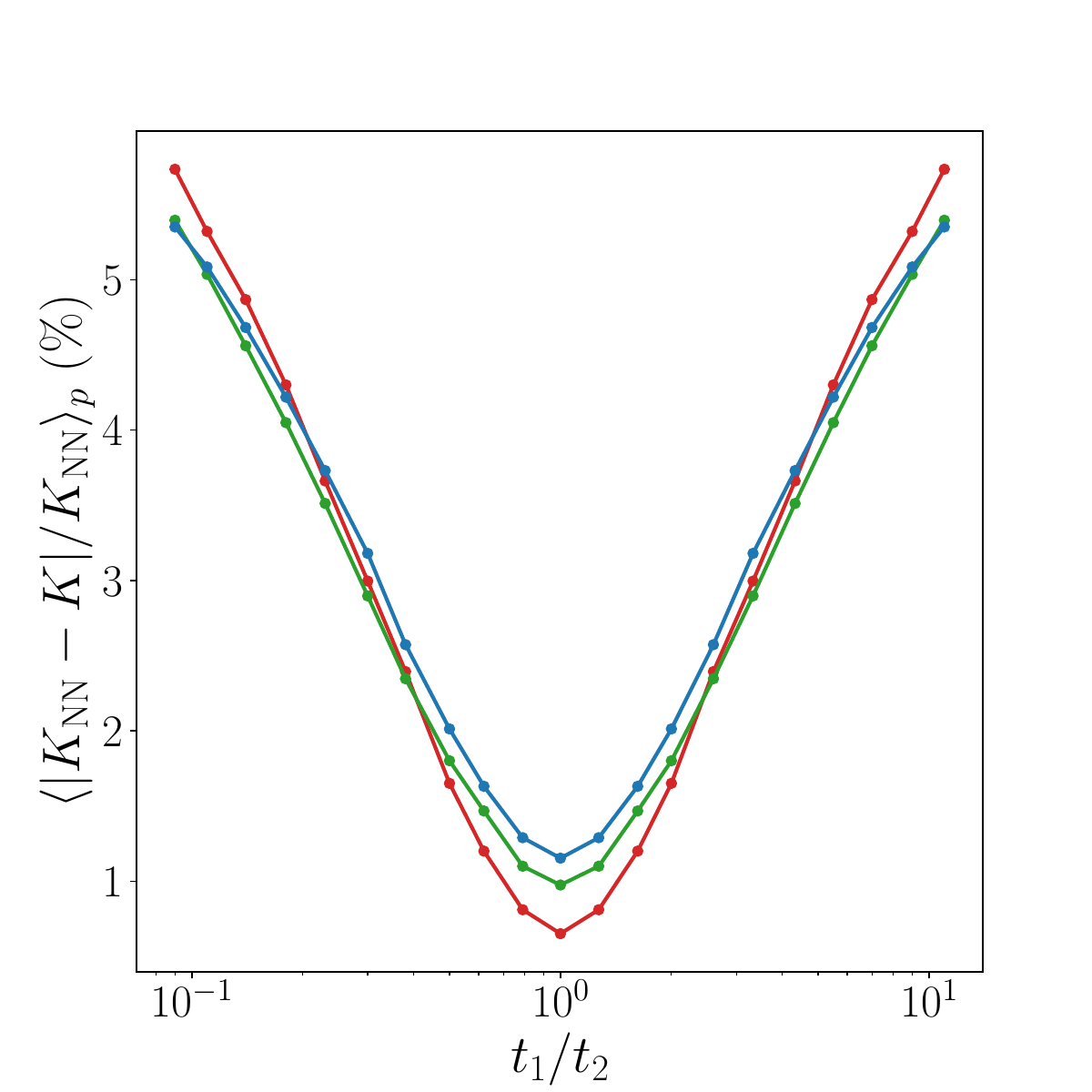}
    \end{minipage}

    \caption{Left: Comparison of the $\sigma$-loss computed from the analytical expression obtained via symbolic regression (triangles) and that obtained by the neural network after training (circles), for all considered values of the ratio $t_1/t_2$. Results are shown for three different values of the complex structure parameter. Right: Percentage deviation $\langle |K_{\rm NN}-K|/K_{\rm NN}\rangle$, averaged over $N$ sampled points on the manifold, as a function of $t_1/t_2$ for the same values of the complex structure parameter. Averaged over the moduli, the $\sigma$-loss and the deviation of the potential are $0.06$ and $3.2\%$, respectively.}
    \label{fig:sigma_loss_and_deviation}
\end{figure}

For the numerical calulation of Ricci-flat metrics we consider bi-cubics defined as zero loci of the polynomial $P$ in Eq.~\eqref{eq:def_P} for the  moduli space points
\begin{equation}
\left\{
(\psi, t_1, t_2) \,\middle|\,
\begin{aligned}
& \psi = m + i n\, , \text{with } m,n = -4, \dots ,5 \, , \text{and } \psi \neq 0,\pm 1 \\
& V(t_1, t_2) = 40\,, ~\frac{t_1}{t_2} {=} \tiny \frac{1}{11}, \frac{1}{9}, \frac{1}{7}, \frac{2}{11}, \frac{3}{13}, \frac{3}{10}, \frac{5}{13}, \frac{1}{2}, \frac{8}{13}, \frac{11}{14}, 1, \dots \, 
\end{aligned}
\right\}
\label{t12val}
\end{equation}
where, in the case of the K\"ahler moduli, the dots indicate the inclusion of all the reciprocals. These constitutes a total of $97 \times 21 = 2037$ points in the full moduli space. Note, we have excluded the values of $\psi$ where the manifold becomes singular. For each of these points in moduli space we sample $N=200{,}000$ points on the corresponding bi-cubic, using the \cymetric~point generator, and then learn the Ricci-flat K\"ahler potential with the \cymetric~$\phi$-model. Fitting to this data, the coefficients $\alpha_i(\psi,t_{12})$ in the Ansatz Eq.~\eqref{eq:bicubicphiAnsatz} are then determined for each of the values $(\psi,t_{12})$ in Eq.~\eqref{t12val}. Finally, symbolic regression is applied to these results in order to find symbolic expressions for $\alpha_i(\psi,t_{12})$.\\[2mm] 
Fig.~\ref{fig:sigma_loss_and_deviation} summarises the accuracy of the construction. In the left panel we show the $\sigma$-loss for both the neural-network metric and the analytic metric obtained via symbolic regression. The $\sigma$-loss measures the extent to which the predicted metric satisfies the Monge--Amp\`ere equation and is defined by
\begin{equation}
\sigma_{\rm loss}
~=~
\frac{1}{N}
\left\|
1-\frac{\det g_{\rm pr}}
{\kappa\,\Omega\wedge\overline{\Omega}}
\right\|_{1},
\end{equation}
where $g_{\rm pr}$ denotes the predicted Ricci-flat metric, $\Omega$ is the holomorphic $(3,0)$ form, $\kappa$ is a moduli-dependent constant, and the $L_1$ norm is evaluated over a sample of $N$ points on the manifold. Thus, $\sigma_{\rm loss}$ corresponds to the average pointwise deviation from the Monge--Amp\`ere equation. In the right panel we show the average percentage deviation of the analytic K\"ahler potential $K$ from its neural-network counterpart $K_{\rm NN}$. Results are displayed for three representative values of the complex-structure modulus. The observed accuracy depends only weakly on this parameter.
\\[2mm]
The $\sigma$-loss of the numerical metric demonstrates that the neural network is able to learn Ricci-flat metrics with good accuracy throughout the range of K\"ahler parameters considered. Although the $\sigma$-loss of the analytic metric is larger than that of the corresponding neural-network solution, it remains sufficiently small for the analytic metric to provide a good approximation to a Ricci-flat metric.
\begin{figure}[h!]
    \centering
    \begin{minipage}{0.45\textwidth}
        \centering
        \includegraphics[width=\linewidth]{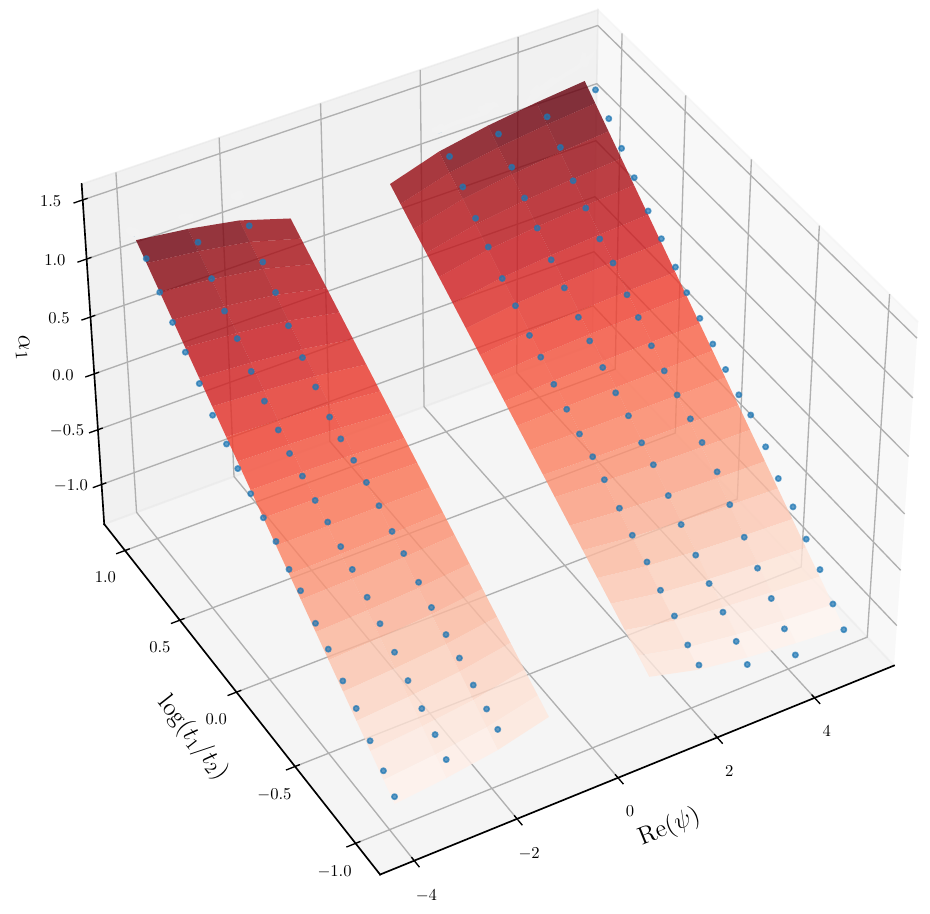}
    \end{minipage}
    \begin{minipage}{0.45\textwidth}
        \centering
        \vspace{0cm}
        \includegraphics[width=\linewidth]{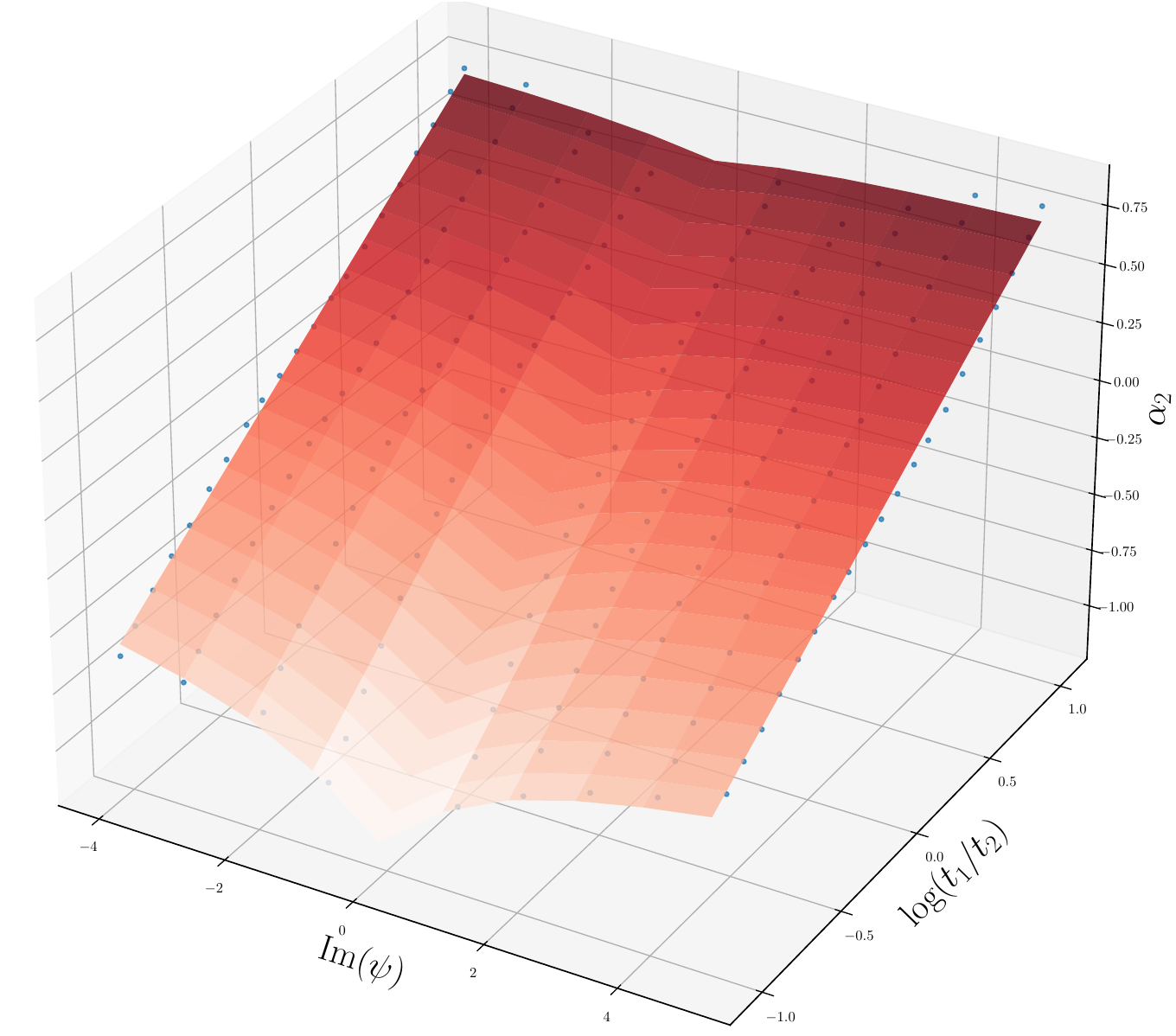}
    \end{minipage}

    \begin{minipage}{0.45\textwidth}
        \centering
        \vspace{0cm}
        \includegraphics[width=\linewidth]{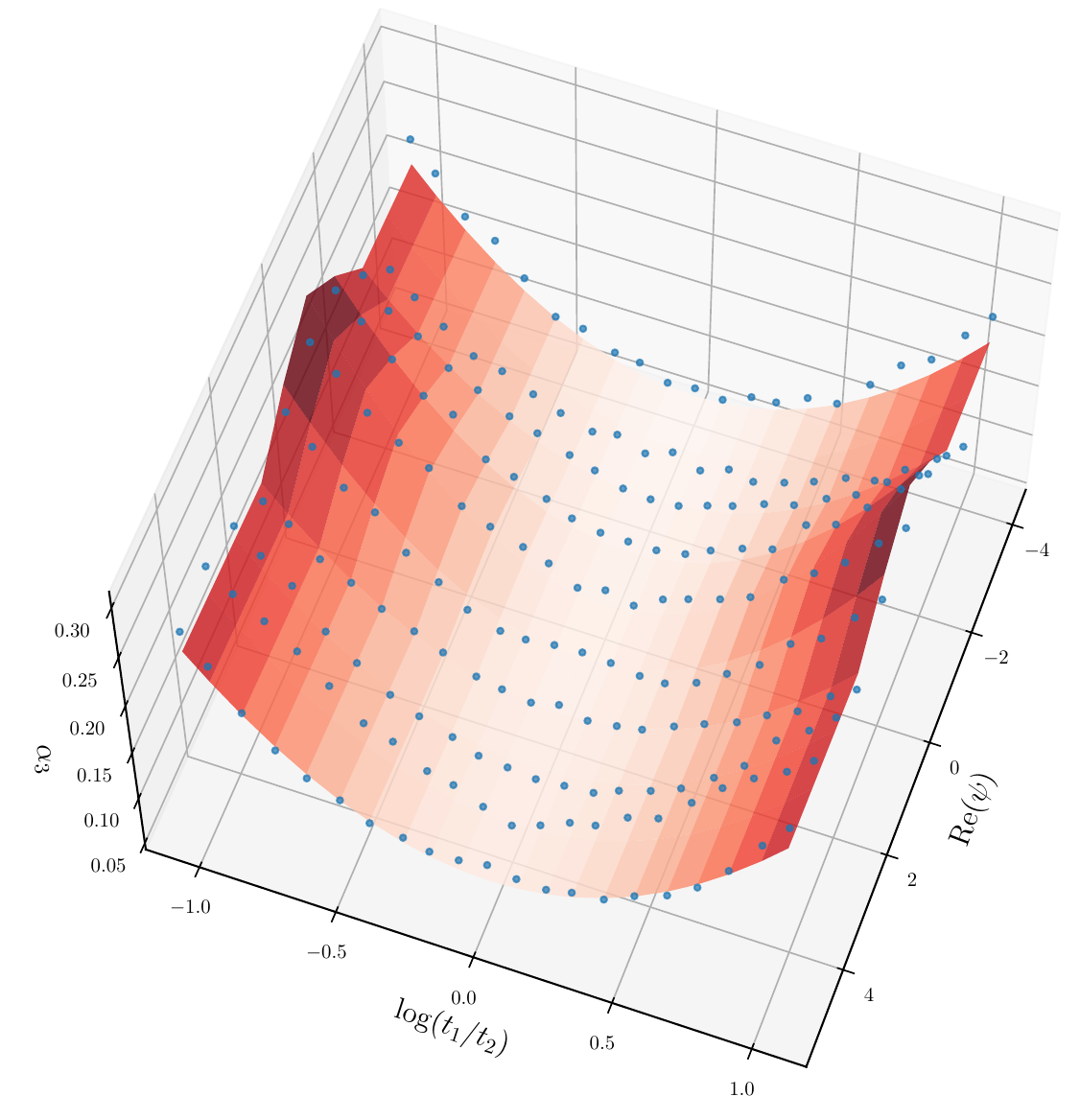}
    \end{minipage}
    \begin{minipage}{0.45\textwidth}
        \centering
        \vspace{0cm}
        \includegraphics[width=\linewidth]{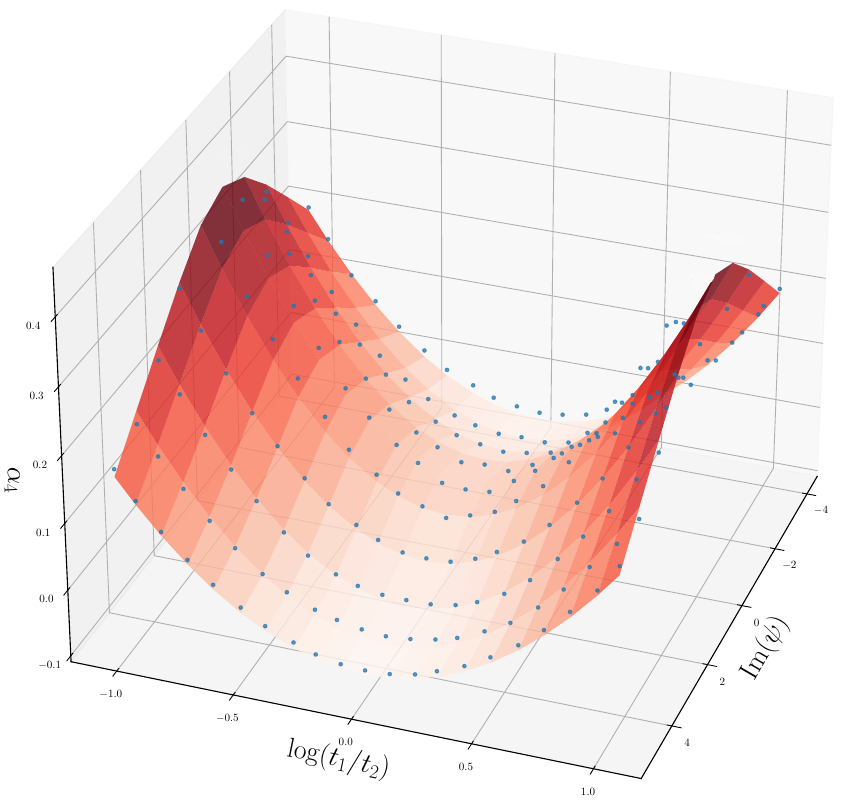}
    \end{minipage}
    \caption{Three-dimensional plots of the first four coefficients $\alpha_i$ as functions of $t_1/t_2$, shown on representative slices of the complex-structure moduli space. The blue points denote the coefficient values obtained from fits to the numerical data, while the red surfaces show the corresponding analytic expressions derived via symbolic regression. In the left panels, $\Im(\psi)$ is fixed to $0$ and $2$ (from top to bottom), whereas in the right panels, $\Re(\psi)$ is fixed to $-1$ and $-2$. The truncation visible in the top-left panel is caused by a singularity at $\psi=0$ in the symbolic-regression expression.}
    \label{fig:3d_plots}
\end{figure}

\vspace{2mm}
\noindent Applying symbolic regression to the numerical values of the coefficients $\alpha_i(\psi,t_{12})$ yields explicit analytic expressions for their dependence on both the complex-structure and K\"ahler moduli. Fig.~\ref{fig:3d_plots} compares these analytic expressions with the corresponding coefficient values extracted from the numerical fits. The analytic results are shown as red surfaces, while the fitted numerical values are represented by blue points. For clarity, the figure displays four representative coefficients and restricts the complex-structure dependence to fixed slices in the $\psi$-plane.\\[2mm]
The top-left panel of Fig.~\ref{fig:3d_plots} shows the results for $\alpha_1$. The plot is truncated at $\Re(\psi)=0$, as the singularity of the manifold at $\psi=0$ is reflected in the corresponding symbolic-regression expression, which becomes singular at that point. Indeed, this singularity at $\psi=0$ appears in almost all the other coefficient expressions. However, despite its severity, the manifold singularity at $\psi=1$ corresponds to a regular point of the symbolic-regression expression for all the coefficients. At present, we do not know of a theoretical reason why singularities of the manifold should necessarily be reflected in the coefficients of the K\"ahler potential. The singularity at $\psi=0$ may therefore be an artefact of the symbolic-regression procedure. Clarifying this issue would require a more detailed analysis based on denser sampling near the singular loci, which lies beyond the scope of the present work.\footnote{Of the two singular members with $\psi=0$ and $\psi=1$ that exhibit sever singularities, the $\psi=1$ case is already in the standard form, degenerating into the union of two smooth non-CY components intersecting normally along a surface. Similar normal-crossing degenerations arise for the one-parameter families of quintics and bi-cubics studied in Ref.~\cite{Lee:2025pue}, in the large-complex-structure limits thereof; note that, for the members with fairly large $\psi$ values of those two families, the analytic approximation based on the $\phi$-model was observed to reproduce the numerically determined Ricci-flat K\"ahler potential to within a few percent.}
\\[2mm]
\begin{figure}[h!]
    \centering
    \includegraphics[width=0.3\textwidth]{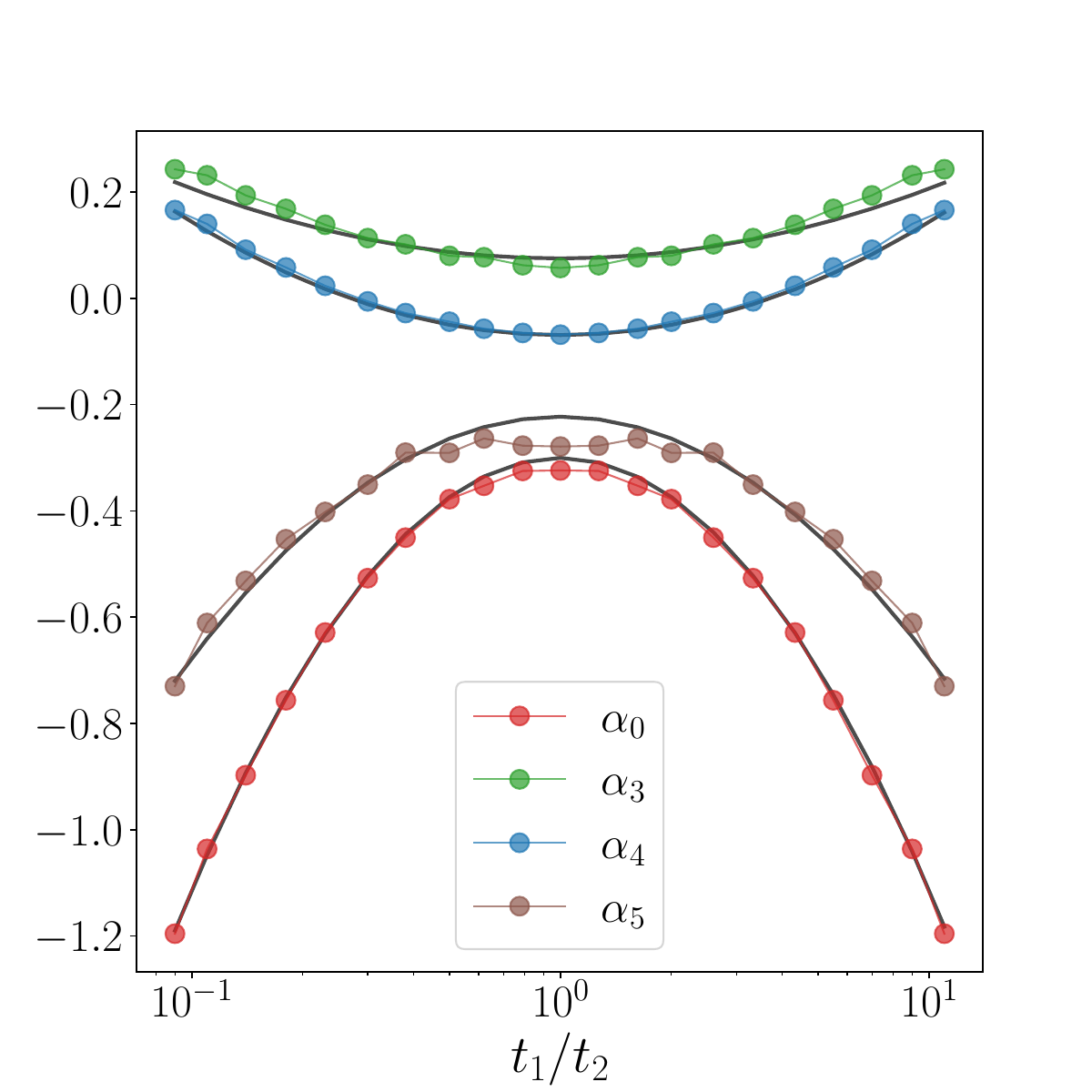}
    \includegraphics[width=0.3\textwidth]{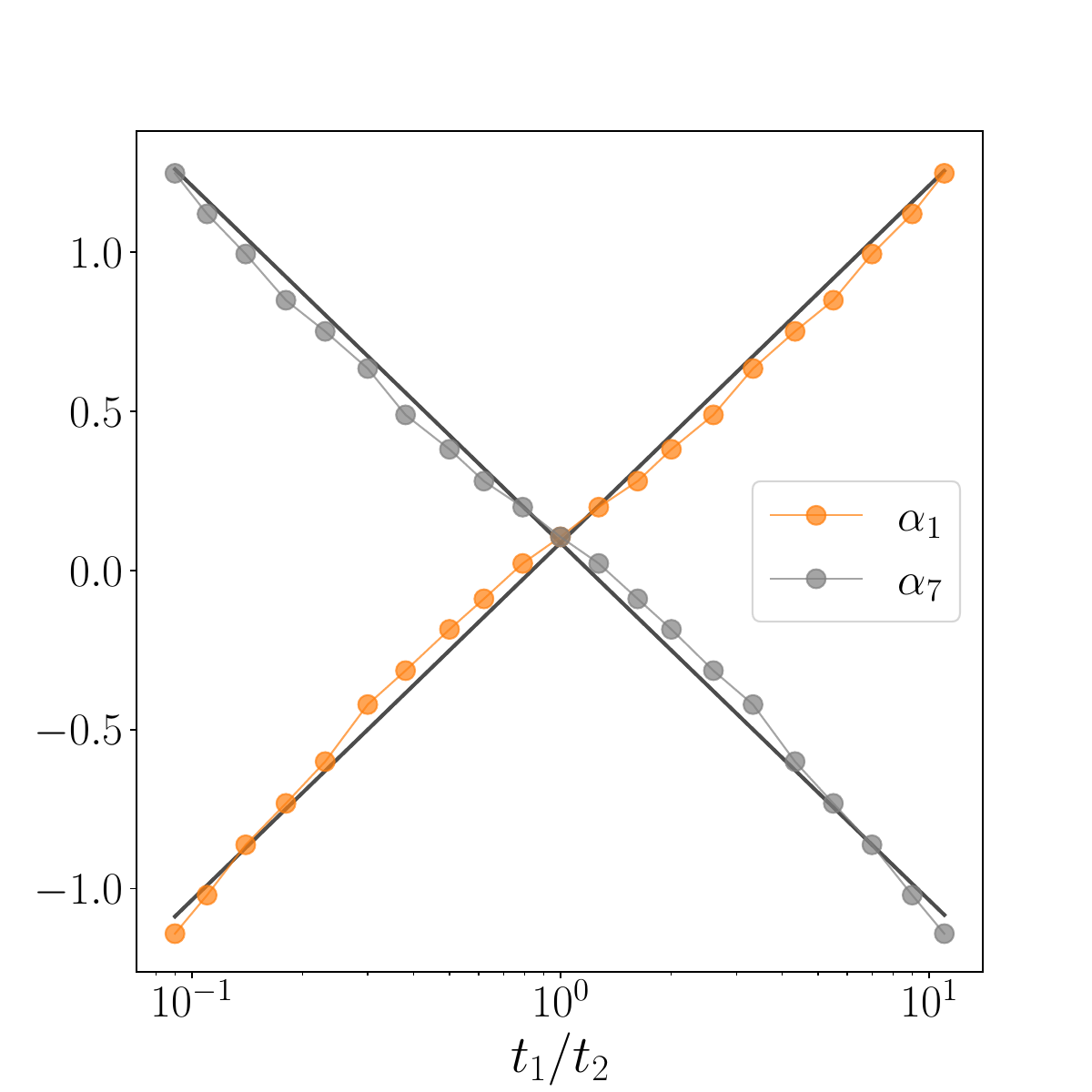}
    \includegraphics[width=0.3\textwidth]{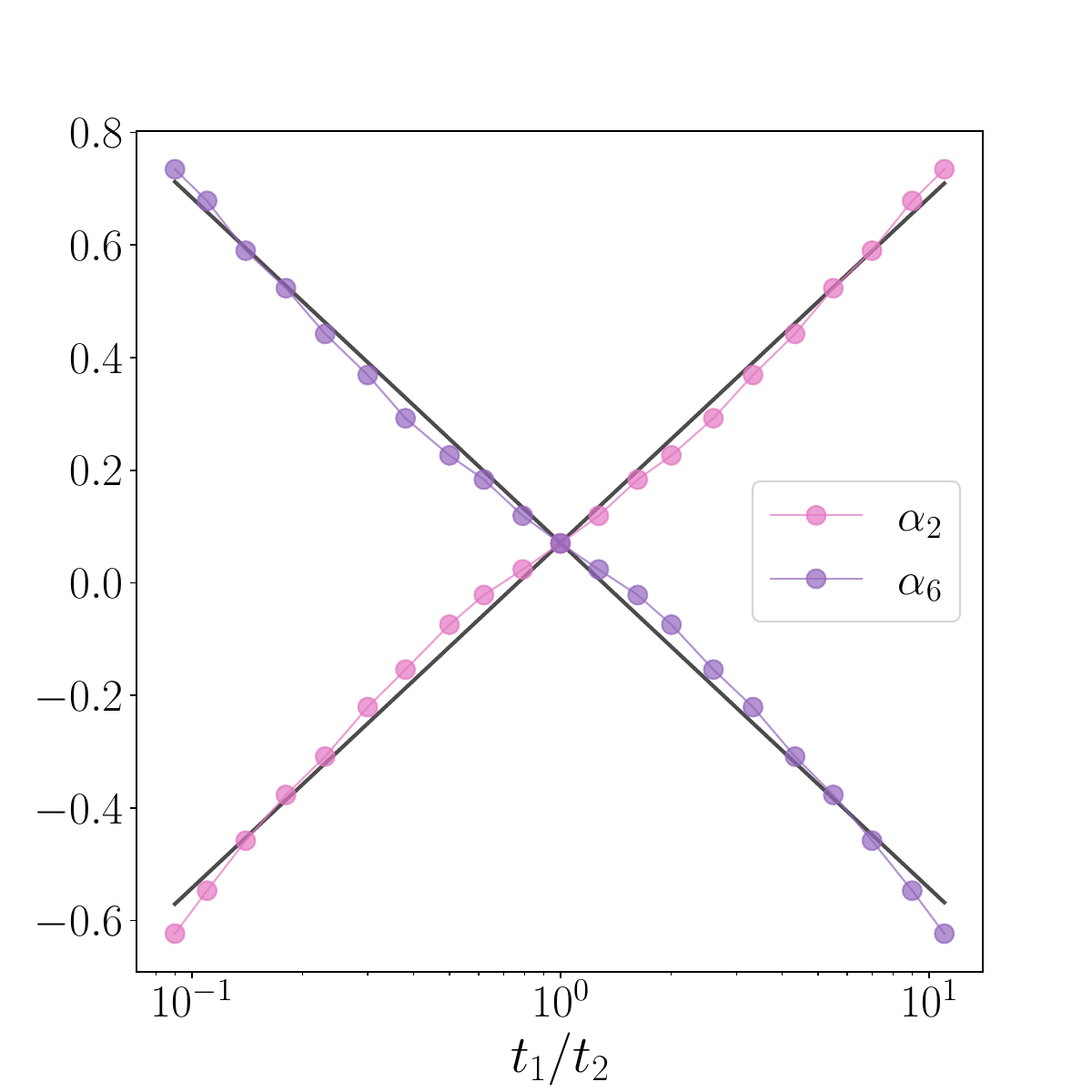}
    \caption{Best-fit numerical values of the coefficients $\alpha_i$ as functions of the moduli ratio $t_1/t_2$ for fixed values of $\psi$.  The corresponding analytical expression obtained using symbolic regression is plotted as a black line.}
\label{fig:alpha_plots_t12}
\end{figure} 

\noindent The lower panels of Fig.~\ref{fig:3d_plots} already suggest that $\alpha_3$ and $\alpha_4$ satisfy the symmetry properties implied by Eq.~\eqref{eq:alphasymmetry}. These relations become more apparent in Fig.~\ref{fig:alpha_plots_t12}, where the fitted coefficient values are shown as functions of $t_1/t_2$ for fixed $\psi$. The corresponding analytic expressions obtained from symbolic regression are shown as black curves.
Substituting these analytic expressions into Eq.~\eqref{eq:bicubicphiAnsatz} yields the following explicit analytic approximation to the Ricci-flat K\"ahler potential, with full dependence on both the complex-structure and K\"ahler moduli, as written in Eq.~\eqref{eq:bicubicphiAnsatzExplicit}.
\vspace{0.5cm}
\begin{equation}
{\small
\begin{aligned}
& K(x,\psi,t_1,t_2)
~=~K_{\rm FS}\big|_{X} ~+~ 
\frac{V(t_1,t_2)^{1/3}}
{\bigl(\sum_A |x_A|^2\bigr)^2
 \bigl(\sum_A |y_A|^2\bigr)^2}
\Bigg\{
\\[4mm]
&\qquad  
\Big[-0.3
-0.07\,\log\!\bigl(0.38\,|\psi|^2\bigr)
\log^2(t_{12})\Big]
\sum_{a<b}|x_a|^2|x_b|^2|y_a|^2|y_b|^2
\\[1mm]
&\qquad +
(0.38+0.037\,\log|\psi|^2)
(0.18+\log t_{12})
\sum_{\substack{a\neq b,c\\ b<c}}
|x_a|^4|y_b|^2|y_c|^2
\\
&\qquad 
+
(t_{12} \to t_{21}, \, x \leftrightarrow y)
\\[4mm]
&\qquad 
+
\Big[
\log\!\bigl(|\psi|^2+0.85\,\Re\psi\bigr)
\bigl(0.064-0.002\,\log t_{12}\bigr)
+0.32\,\log t_{12} -0.1
\Big] \sum_{a\neq b}|x_a|^4|y_a|^2|y_b|^2
\\
&\qquad  + (t_{12} \to t_{21}, \, x \leftrightarrow y)
\\[4mm]
&\qquad 
+
\Big[
0.08
+\bigl(0.024+0.014\,e^{-0.564(\Re\psi)^4}\bigr)
\log^2(t_{12})
\Big]
\sum_{a\neq b}|x_a|^4|y_b|^4
\\[4mm]
&\qquad 
+
\Big[
\bigl(0.07-0.001(|\psi|^2-0.28\,\Im\psi)\bigr)
\bigl(1.5-\log|\psi|^2+\log^2(t_{12})\bigr)
\Big]
\sum_a |x_a|^4|y_a|^4
\\[4mm]
&\qquad 
-
\Big[
0.22+0.02\,\log\!\left((\Im\psi)^2+3.71(\Re\psi)^2\right)\left(e^{-0.46 \Re\psi + 0.13 (\Re\psi)^2}+\log^2 (t_{12})\right)
\Big] \sum_{\substack{a,b,c\,\mathrm{distinct}}}
|x_a|^2|y_a|^2|x_b|^2|y_c|^2
\Bigg\} \, 
\end{aligned}
}
\label{eq:bicubicphiAnsatzExplicit}
\end{equation}
As before, we note that the Fubini--Study contribution is restricted to the CY manifold. As a result, it acquires an implicit dependence on the complex-structure parameter $\psi$ through the defining polynomial from Eq.~\eqref{eq:def_P}. As stated above, the associated $\sigma$-loss and average percentage deviation from the numerical K\"ahler potential are $0.06$ and $3.2\%$, respectively. We emphasise that this level of accuracy is achieved already for $k=(k_1,k_2)=(2,2)$, corresponding to a relatively low-degree line bundle. A noteworthy outcome of our analysis is that accurate analytic approximations do not require large values of $k$; the essential geometric features are already captured by a comparatively simple Ansatz.

\section{Conclusions and outlook}
In this paper, we have developed a method for constructing approximate Ricci-flat K\"ahler potentials with explicit dependence on both complex-structure and K\"ahler moduli. The construction combines numerical Ricci-flat metric data with a suitable analytic Ansatz for the K\"ahler potential, based on sections of a line bundle $L={\cal O}_X(k)$, together with symbolic regression. We have implemented this approach for a one-parameter family of bi-cubic CY hypersurfaces in $\mathbb{P}^2\times\mathbb{P}^2$, but the method is sufficiently general to be applied to a broad class of CY manifolds, including hypersurfaces and complete intersections in toric ambient spaces.
\\[2mm]
Although our results only provide approximately Ricci-flat metrics we still anticipate meaningful applications in the context of \CY string compactifications. As mentioned, in such compactifications, the physically relevant quantities depend on both complex-structure and K\"ahler parameters. Having metrics available which display the dependence on these moduli explicitly, allows for `tracking' the value of physical quantities across the moduli space. The approximate nature of our results is not necessarily prohibitive in this context. Most string calculations, certainly at the level of physical models, are carried out using various approximations, including the $\alpha'$ and loop expansions of string theory. The approximate nature of our results simply provides another contribution to the inevitable error in such calculations. As to the size of this error, it is instructive to note that the calculation of Yukawa couplings performed in Ref.~\cite{Constantin:2024yxh} shows that using a Fubini-Study metric (that is, the first term in the Ansatz~\eqref{eq:general_ansatz} only) leads to quark and lepton masses that deviate by only about $20\%$ from those obtained with a machine-learned Ricci-flat metric. Our result is a very significant improvement on merely considering the Fubini-Study metric which suggests it may lead Yukawa couplings with percent level errors.\\[2mm]
There are various extensions of the present work. First, one might attempt to carry out the programme described in this paper for more complicated \CY manifolds and/or higher-dimensional moduli spaces.
There are several versions of this goal. The most conservative is to remain within highly symmetric families and enlarge the parameter space gradually, adding one modulus at a time. A more ambitious version would treat several complex-structure and several K\"ahler moduli simultaneously, perhaps with sparse sampling combined with stronger inductive biases at the symbolic-regression stage.\\[2mm] 
In view of calculating physical quantities it seems desirable to extend the present philosophy to the additional geometric ingredients entering such calculations. For example, it might well be possible to obtain approximate analytic results for Hermitian Yang--Mills connections on bundles, and harmonic forms.\\[2mm]
An important direction for future work is to determine how large the line bundle degree must be in order for the analytic metric to achieve a level of Ricci-flatness comparable to that of the numerical solution. While the analytic metric already provides a good approximation to a Ricci-flat metric, it remains an open question whether increasing the degree of the associated line bundle can systematically reduce the remaining discrepancy. \\[2mm]
Finally, a natural question is why one should employ a neural network to learn the Ricci-flat K\"ahler potential when, in principle, the analytic metric itself could be learned directly using symbolic regression with a loss function enforcing Ricci-flatness. While this certainly represents an interesting direction for future investigation, there are indications that such an approach may be computationally prohibitive~\cite{Mirjanic:2024gek}.

\section*{Acknowledgements}
AC~and LAN~are supported by the Royal Society grant DHF/R1/231142. AL is supported by the STFC consolidated grant ST/X000761/1. The work of SJL is supported by the Yonsei University Research Fund 2026-22-0183.

\printbibliography
\end{document}